\documentclass[3p,twocolumn]{elsarticle}
\usepackage{graphicx}

\usepackage{eurosym}
\usepackage{siunitx}
\DeclareSIUnit{\dB}{dB}
\DeclareSIUnit{\EUR}{\euro}
\usepackage{soul}
\usepackage{flushend}

\usepackage{helvet}

\fontfamily{phv}\selectfont 

\usepackage{wrapfig}
\usepackage{xcolor,colortbl}
\usepackage{verbatim}

\definecolor{dbl}{rgb}{0.61,0.69,0.79}
\definecolor{lbl}{rgb}{0.65,0.76,0.95}
\definecolor{lrd}{rgb}{0.97,0.79,0.67}
\definecolor{grn}{rgb}{0.66,0.81,0.55}
\definecolor{lgr}{rgb}{0.9,0.9,0.9}
\definecolor{dgr}{rgb}{0.8,0.8,0.8}
\definecolor{wht}{rgb}{1,1,1}

\usepackage{xcolor}

\title{''Small is beautiful'' in NMR}
\author{Jan G. Korvink, Neil MacKinnon, Vlad Badilita, Mazin Jouda}
\address{Institute of Microstructure Technology, Karlsruhe Institute of Technology, P.O.Box 3640, 76021 Karlsruhe, Germany }

\begin{document}

\begin{abstract}
In this prospective paper we consider the opportunities and challenges of miniaturized nuclear magnetic resonance. As the title suggests, (irreverently borrowing from E.F. Schumacher's famous book), miniaturized NMR will feature a few small windows of opportunity for the analyst. We look at what these are, speculate on some open opportunities, but also comment on the challenges to progress. 
\end{abstract}

\maketitle

\section{Introduction}
The miniaturization of devices belongs to a research field known as MEMS, short for micro-electro-mechanical systems. The field was born out of a 1950's talk by Richard Feynman, \textit{''There's plenty of room at the bottom''} that inspired engineering scientists of the time and still does to this day. The MEMS community collectively develops microscopically sized detectors and actuators, combined with electronic systems, sometimes integrated with complex microfluidics for handling minute amounts of sample called Lab-on-a-Chip (LOC) systems. All sorts of applications are targeted, covering biology, chemistry, and even chemical engineering, in addition to those known from consumer products such as the instruments found in smart devices. During the time the first author was a PhD student (early 1990's), our professors were seeking the ''killer applications'' for MEMS, hoping that their field would establish itself before running out of steam. Their wish was granted: automobiles and smart phones created the necessary initial market pull, with the productive research field giving rise to billions of dollars worth of business over the course of a few decades. Miniaturized detector systems are clearly here to stay.

MEMS systems are currently revolutionizing measurement science, yielding more sensitivity and information, with less effort, in the smallest of space. Thus electron microscopy, atomic force microscopy, plasmonics, optical spectroscopy, and so on, all benefit significantly from MEMS technology, opening up fantastic new opportunities. It is therefore well worth considering whether NMR and MRI could equally benefit from this approach. For this we need to look more deeply into the crystal ball, to see whether we can find pathways that lead to some ''killer applications''.

\begin{figure}[t]
\label{fig:TheRoadToCompaction}
\centering
\includegraphics[width=\columnwidth
]{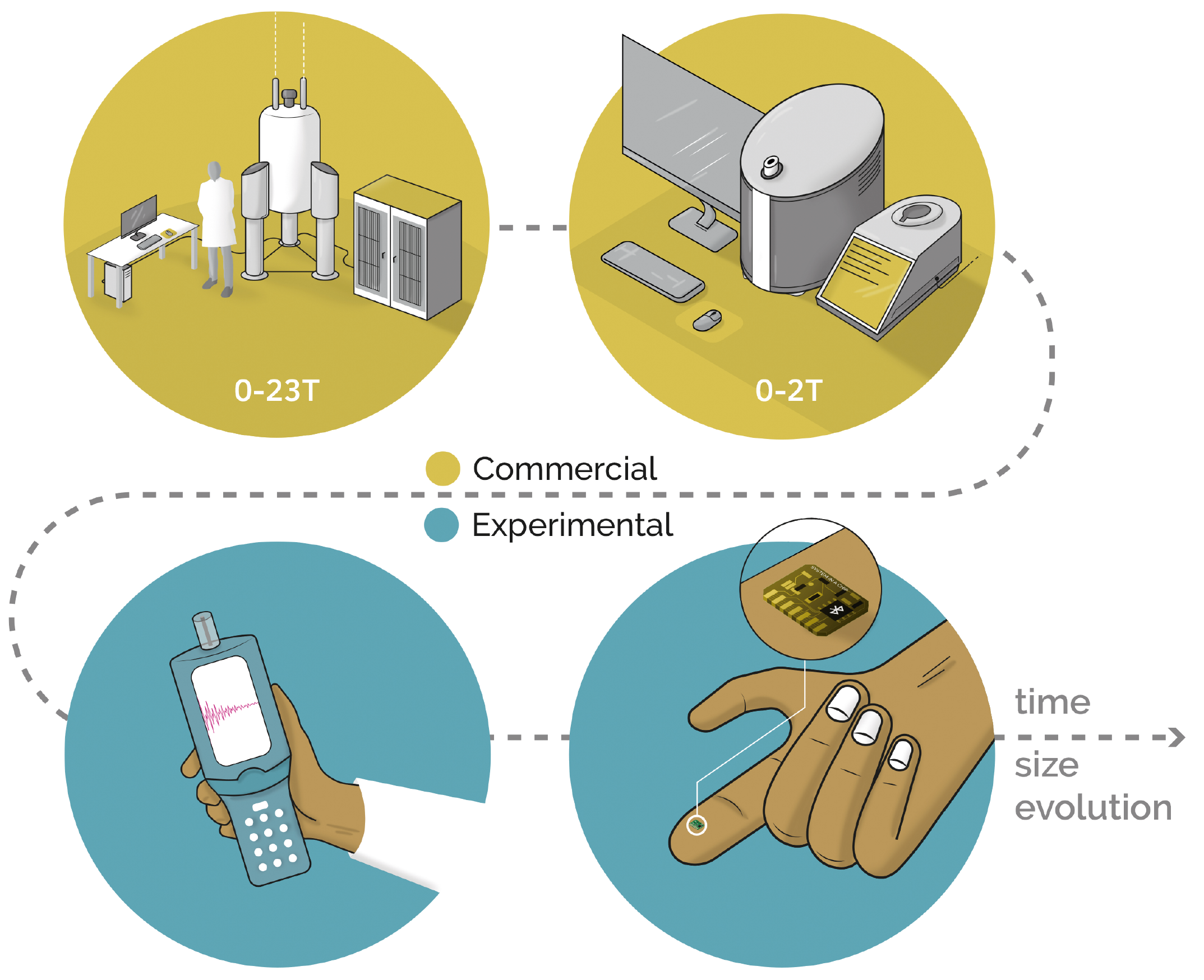}
\caption{The entire NMR system is becoming available in more compact format, facilitated by the shrinking of electronics, magnets, and resonators. However, all but the large laboratory NMR systems have a field strength on the order of 1 Tesla.  On the other hand, compact palmtop and chip-scale NMR systems have yet to make it to generally available products.}
\end{figure}

\section{Why bother with small?}
Since the mid 1990's, numerous publications \cite{ Olson:1995fr,Massin:2002wu,Rugar:2004bc,Bentum2007,Sakellariou2007,Maguire:2007ko,Kratt2010,Danieli:2010ep,Anders:2011if,Ryan2012,WRACHTRUP2016225,Anders:2018} have reported on efforts to miniaturize NMR. Many of these reports are truly impressive, either because of the increased NMR performance, or in the sophistication of the technical systems. To date, very few of these developments have made it into products (see Figure~\ref{fig:TheRoadToCompaction}), so that accessibility is limited to only a few practitioners and their associates. Overall, such a progression is not unusual for any field of endeavour, but of course there comes a time when it is prudent to ask: \textit{''why bother?''} Once the novelty of \textit{''because we can''} is over, one might hope for either more science, or more applications, in order to justify further exploration and adoption.

The core of an NMR system is the detector. In an elegant paper, John Sidles showed that the signal-to-noise ratio (SNR) of a miniaturized LC resonator and a magnetic resonance force microscope (MRFM) are of the same order of magnitude, even though the techniques have very different absolute spin sensitivities \cite{PhysRevLett.70.3506}. 
\textcolor{black}{Sidles' paper showed also that it is possible for magnetic resonance force microscopy to outperform inductive NMR, provided that a proper mechanical resonator with small mass, high resonance frequency, and high quality factor is used. However, the complexity, particularly with liquid-state samples, and limited applicability of this technique, together with access to the toolbox of sophisticated modern NMR experiments, are the reasons why inductive detection is, and is most likely to stay, overwhelmingly used.} Since this covers our length-scale of interest, we will therefore restrict our attention to inductive measurements using coils (also covering striplines), knowing that other approaches such as the use of nitrogen vacancies in diamond, or cantilever probes, may offer slightly better sensitivities for specific situations.

In the NMR experiment, the signal-to-noise ratio (SNR) per sample volume, measured at the terminals of the RF coil, can be described by~\cite{hoult1976signal}
\begin{equation}
    \frac{\textrm{SNR}}{V_{s}}=\frac{KN\gamma\hbar^2I(I+1)/(3k_BT_s)\cdot\omega_0^2 \cdot B_1}{\sqrt{4k_BT_c\Delta f}\cdot\sqrt{R_\textrm{noise}}},
\end{equation}
where $K$ is a field inhomogeneity factor, $N$ is the number of resonant spins per unit volume, $\gamma$ is the gyromagnetic ratio, $\omega_0$ is the Larmor frequency, $\hbar$ is the reduced Planck's constant, $I$ is the spin quantum number, $k_B$ is the Boltzmann's constant, $T_s$ is the sample's temperature, $T_c$ is the coil's temperature, $\Delta f$ is the measurement bandwidth, $B_1$ is the magnetic flux density per unit current flowing in the coil, and $R_\textrm{noise}$ is the AC resistance including both the sample and the coil losses. In microcoils, sample losses can be neglected since the coil is the major source of noise.
\textcolor{black}{This assumption is valid only for non-conductive samples, or samples with very low concentration of salt, in which case the effective resistance that represents sample losses is too small and thus negligible. However, if the salt concentration in a sample increases, then the inductive losses within the sample increase and as a result, the overall coil sensitivity decreases \cite{hoult1979sensitivity}. This effect becomes more pronounced with coils that have low noise contribution, such as the cryoprobes \cite{kelly2002low}. Thus, defining a size of the coil below which sample losses can be neglected, depends mainly on the noise performance of the coil, as well as the conductivity of the sample.}

\textcolor{black}{Neglecting sample losses for simplicity, } the SNR per sample volume of a microcoil is then proportional to the square of the frequency $\omega_0^2$, and the coil sensitivity $B_1/\sqrt{R_\textrm{coil}}$:
\begin{equation}
    \label{eq:SNR_NMR}
    \frac{\textrm{SNR}}{V_s}\propto \frac{\omega_0^2\cdot B_1}{\sqrt{R_\textrm{coil}}}.
\end{equation}
$R_\textrm{coil}$ is generally dependent on the coil geometry, the wire used in its construction, and by extension, the frequency of operation.  We will consider a solenoid as an example to demonstrate the benefits of miniaturization when accounting for proximity and skin effects, although the discussion can be generalized to any coil geometry.  The magnetic flux density per unit current of a single layer solenoid of multiple windings $n\gg 1$, with radius $r_\textrm{coil}$, and length $l$, can be calculated along the axis of the coil by the direct application of the Biot-Savart law \cite{kraus1973electromagnetics}: 
\begin{equation}
    \label{eq:B_solenoid}
    B_1=\frac{\mu_{s} n}{r_\textrm{coil}\sqrt{4+(l/r_\textrm{coil})^2}},
\end{equation}
where $\mu_s$ is the sample's permeability. The AC resistance of the solenoid, assuming the radius of the wire to be significantly larger than the skin depth ($r_\textrm{wire}\gg \delta$), can be approximated as \cite{cho1988nuclear,mcfarland1992three,hoult1979sensitivity}:
\begin{equation}
    \label{eq:R_solenoid1}
    R_{\textrm{coil}} \approx \frac{3\rho n^2 r_\textrm{coil}\xi}{ l \delta},
\end{equation}
where $\rho$ is the resistivity of the coil material, $\xi$ is a proximity factor that depends on the number of coil layers and windings, and $\delta=\sqrt{2\rho/(\mu_\textrm{wire} \omega_0)}$ is the (frequency dependent) skin depth. Substituting Equations~\ref{eq:B_solenoid} and \ref{eq:R_solenoid1} into Equation~\ref{eq:SNR_NMR}, and assuming that the number of windings and the ratio $l/r_\textrm{coil}$ are maintained constant, results in:
\begin{equation}
    \label{eq:SNR_NMR1}
    \frac{\textrm{SNR}}{V_s}\propto \frac{\omega_0^{7/4}}{{r_\textrm{coil}}}.
\end{equation}
Thus, if the coil radius is reduced by a factor of 2, then the coil sensitivity as well as the SNR per sample volume doubles, whereas the excitation power required for a certain flip angle decreases by a factor of 4.

It is therefore instructive to briefly consider the scaling laws for detectors. We assume a flat circular coil (saddle or solenoidal coils will behave similarly) with diameter $D$ and wire diameter $d$ (important for the resistance), and a linear miniaturization scaling factor $0<\alpha_L\leq 1$ so that $D_\textrm{new}=\alpha_L\cdot D_\textrm{old}$:
\begin{itemize}
    \item \textbf{Volume and signal:} The sample volume and hence NMR signal strength scales as $\alpha_V=\alpha_L^3$. One clearly accesses a rapidly decreasing number of spins by going down in size.
    \item \textbf{Radiofrequency resistance:} Since $R=\rho l/(\pi d\delta)$, with $\rho$ the resistivity, $l$ the wire length, and $\delta$ the skin depth, the coil resistance is a constant under scaling as long as $\delta\ll d$.
    \item \textbf{Bandwidth and excitation field strength:} For a given current $I$, the $B_1$-field scales as $\alpha_B=\alpha_L^{-1}$, as does the excitation bandwidth. See below for more detail.
    \item \textbf{Power:} The RF power required to drive a given current scales as $\alpha_P=\alpha_L^{2}$, as long as sufficient skin depth in the conductor wires is retained, i.e., $\delta\ll d$.
    \item \textbf{Signal-to-noise ratio (SNR):} The SNR scales as $\alpha_\textrm{SNR}=\alpha_L^{-1}$ \cite{Hoult:1976dw}, 
    \textcolor{black}{ provided that the skin depth is significantly smaller than the wire diameter, $\delta \ll d$}.
     \item \textbf{MAS rotation speed:} For a given material and geometry, the achievable rotation frequency scales with the inverse of the diameter: $\alpha_\textrm{MAS}=\alpha_L^{-1}$.
\end{itemize}

Miniaturization thus brings along some interesting general advantages, such as more excitation bandwidth, less power for a given bandwidth, and potentially, considerably more SNR. It also offers a few further features that are potentially useful for the practitioner:
\begin{itemize}
\item \textbf{Smaller compartments, for example in biology.} The size of the smallest detectable voxel determines the kind of system for which NMR can provide distinguishable data, which is useful especially in biological systems.
\item \textbf{Naturally small samples.} Cells, cell clusters, or tiny organisms such as 
\textit{C. elegans}, are ideal small systems for metabolomic studies. Tissue biopsies or spinal extract from ailing patients yield unique information on disease progression, but must be kept as small as possible to minimise patient discomfort. The bacterial production of bio-molecules is time-consuming, so that smaller samples are a boon because they are quicker to establish. 
\textcolor{black}{Thin films and fibers present functionality dependent on the arrangement of small quantities of material}.  Also, lab-scale chemical synthesis often results in tiny, yet valuable samples. 
\item \textbf{Tight spaces:} Arteries and other biological ducts need MRI at the highest spatial resolution, but have little extra space to maneuver, requiring highly miniaturized solutions. The same is true for samples inside high pressure diamond anvils.
\item \textbf{Less analysis time.} 
\textcolor{black}{Better} SNR naturally leads to quadratically shorter measurement times.
\item \textbf{Improved excitation.} A consequence of the the reciprocity principle ensures that small resonators, equipped with lower inductance than their macroscopic counterparts, create higher fields per unit current, and thus excite spins over a broader range of frequencies.
\item \textbf{Better use of magnetization real estate.} Miniaturization offers the opportunity to advantageously parallelize NMR, by placing many NMR experiments side-by-side. This will not only reduce the cost-of-ownership, but will also lead to dramatically more SNR, and more freedom in experimentation.
\item \textbf{More efficient signal processing.} NMR miniaturization meets and benefits from the well-established field of CMOS technology. When NMR signal processing chains are implemented in CMOS chips, all of the advantages of integration come to play: higher processing speed, less power, channel multiplexing, shorter cables reducing losses, to name the most obvious. 
\item \textbf{Ubiquity.} The level of expertise needed for NMR is currently extraordinarily high. Miniaturized systems can help drive NMR into novel application areas and simplify modes of usage.
\item \textbf{Hyphenation and correlation.} It is becoming increasingly easy to bring extra technology 
into the bore of the magnet, including novel ''excitors'', ''detectors'', and ''influencors''. 
Such tiny MEMS systems open up the door for new kinds of correlative measurements, expanding the conclusions we can draw from an NMR experiment.
\item \textbf{Complex sample handling.}  Microfluidic systems that are compatible with small detector platforms open the door to the lab-on-a-chip philosophy of sample management.
\item \textbf{More efficient hyperpolarization:} At the microscale, less transport distance between hyperpolarization and detector leads to less signal loss \cite{C8LC01259H}.
\item \textbf{Faster MAS.} In magic angle sample spinning, smaller samples would yield faster spinning. The current rotation speed limit for MAS is \SI{150}{\kilo\hertz}, and for reasons of material strength requires rotors of around \SI{0.5}{\milli\metre} diameter, close to the limit of conventional manufacturing precision. 
\textcolor{black}{Rotor geometry has recently been shown to be a degree of freedom \cite{Chen_Magic_2018}. However, there are also potential challenges associated with MAS miniaturization: whereas spheres appear to be more stable than cylinders, the wall thickness requirement of small rotors depends critically on the ultimate strength to density ratio of the wall material, and adding more wall thickness is at the cost of sample volume.}
\end{itemize}

Before proceeding with a discussion on what to miniaturize, it is also instructive to consider the choices and compromises we have to make. When an NMR signal is detected from a volume of matter, we either try to maintain conditions as constant as possible over the volume, thereby treating the entire sample as an ''ensemble'' over which we average, or we use some scheme to select a subset of the sample, for example by applying a gradient field that locally varies the magnetic resonance condition, as is routine in MRI. When a detector is miniaturized with respect to the sample, both its useful interaction volume and its penetration depth decrease, which has consequences for the spatial localization of spins, and for the number of spins in the observed ensemble. Furthermore, there is no point in using a detector beyond its intrinsic sensitivity limit, which fundamentally restricts the ensemble size. This in turn limits the concentration of detectable spins, for example in biological systems \cite{Badilita:2012fc}.

The ideal choice for a particular detector type (microcoil, stripline, magnetic resonance force microscope, NV centre, etc.) is to be well matched to the geometry of the sample and to the kind of questions we wish to answer. The most accommodating sample to deal with is a liquid, fitting into any space available. The least accommodating is a live cell, being mechanically fragile, and requiring a cosy environment and life-sustaining consumables, including the management of waste products \cite{C3RA43758B}.

\section{What can we miniaturize?}
Both NMR and MRI are instrumentation-intensive activities. Apart from solid state magic angle spinning and benchtop NMR magnets, miniaturization is certainly not yet (2019) part of the established commercial equipment roadmap, as was sketched in Figure~\ref{fig:TheRoadToCompaction}. Desirable NMR and MRI magnets are very large and sensitive instruments, and so are well-equipped consoles (our console includes gradient amplifiers, an MAS control unit, and 6 RF channels, running over three 19'' cabinets, each 1.5 m high and 0.5 m wide). Therefore we might first consider what miniaturization could offer for the existing equipment base, and what the consequences could be.

\begin{figure}[t]
\centering
\includegraphics[width=\columnwidth
]{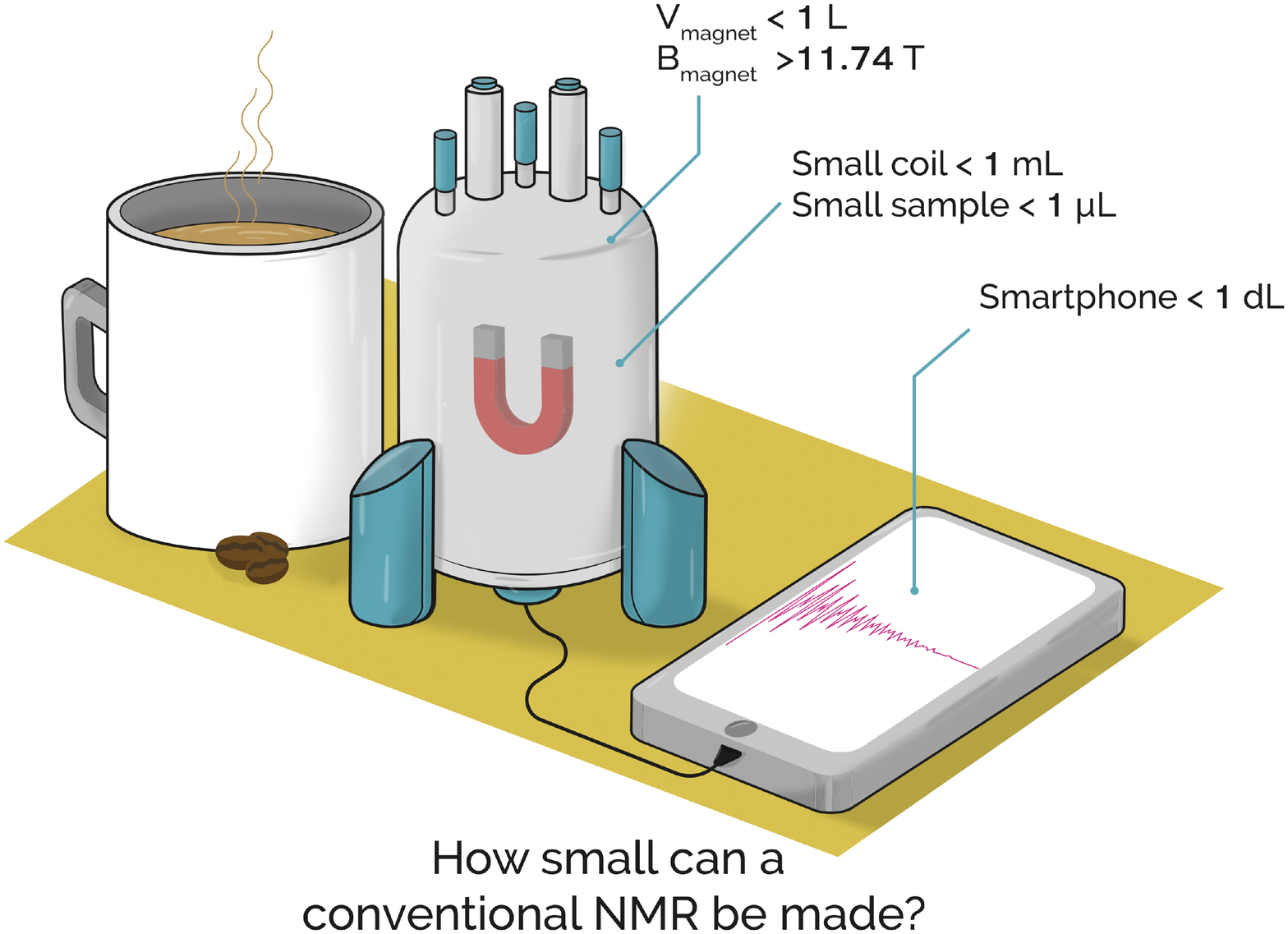}
\caption{A compact NMR system could be achieved by a regular shrinking of each of its parts. Only the miniaturized superconducting high field magnet is not yet on the market. }
\label{fig:AllSmallHighField}
\end{figure}

The NMR magnet is probably the most obvious candidate for miniaturization. A number of vendors have demonstrated impressive tabletop sized permanent magnet systems, with magnetic fields reaching up to \SI{2}{\tesla}, which is probably near to the upper limit in field strength for classical Halbach design magnets. Already now, these benchtop systems are opening up NMR to a much larger and broader community. Two aspects will almost certainly be game changers. For one, the price must still come down -- at a target of 20~k\euro\ most laboratories could probably afford a magnet. For two, the field must go up. Ever since the advent of closed cycle cryogen free magnet technology, and high-$T_c$ magnetic wire, there is no ''engineering'' reason left why we cannot achieve a \SI{23}{\tesla} shielded benchtop system (the vision is to have this magnet propped up next to a laptop in the office, as tiny as indicated in Figure~\ref{fig:AllSmallHighField})\cite{yoon201626,Kim_Design_2017}. Of course this is a great idea even at half of that field strength; each student in the lab would have their own instrument, and productivity would go up tremendously. But more than that, we would be able to run different kinds of experiments, and industry could use high field NMR for new purposes. Because the bore would be much smaller, we would have to use smaller detectors and would need less sample, a boon for many applications.

The next candidate for miniaturization is the spectrometer console. Consider this analogy from electronics: when the first author was a PhD student at the ETH-Zurich,  the CRAY-XMP/28 supercomputer was used for his thesis calculations\footnote{The CRAY-XMP/28 computer had 2 processors, 64 MB of RAM, and a clock speed of 100 MHz. It cost about 15 million dollars. Today, a raspberry pie has 512 MB of RAM, runs at 700 MHz, and costs 38 dollars.}. It was a huge box of electronics, with a water cooling room in the basement to allow the circuits to work at full power. His current smartphone has more memory, and more signal processing power, than the Cray-XMP of 1990. And it is air cooled, running off a battery. This radical miniaturization mindset has not yet come to NMR consoles, apart from a number of pioneering research papers \cite{Lee:2008bx,Anders:2011if,Takeda:2007cx}. The immediately accessible technology is perhaps not the smart phone, but rather the so-called Software Defined Radio (SDR) platform \cite{Tang:2011hz}, which is capable of running pulsed NMR experiments. By defining the operating principles using a software specification, a range of SDR hardware platforms have emerged, on which Rx-Tx experiments can be performed by portable software, over frequencies ranging from \SI{20}{\mega\hertz} to \SI{5.4}{\giga\hertz}. These circuit boards cost around 250 \euro\ a piece, making a four channel console about 1000 \euro\ in hardware investment! Indeed, there is no ''engineering'' reason left why the SDR hardware cannot be implemented into a single CMOS chip, even in a multi-channel format, achieving an overall console size similar to that of a smart phone \cite{Anders:2011if}.

Gradient systems also have much to offer from being smaller. By reducing the distance between two parallel wires carrying currents in opposite direction, the gradient field strength goes up proportionally. Numerous advances in topology optimization have made it possible to design microgradient coils that produce high gradient strengths at the lowest possible power dissipation \cite{While:ey}. As the coils get smaller, so their inductance reduces too, making it possible to switch the gradients much faster than their macroscopic counterparts. In turn, high precision current sources are required to ensure precise current-time profiles. Because of the smaller size, also gradient coil power consumption benefits from miniaturization.

Dynamic nuclear polarization is an equipment- heavy discipline -- a practitioner effectively needs a second large magnet, one major barrier to widespread use. Much power is lost between a gyrotron and the sample, so one approach could be to optimize power transfer between source and resonator, and here miniaturization technologies could certainly help to improve waveguides and resonators. Solid state sources based on diodes,  have emerged as viable miniaturization candidates challenging the use of gyrotrons, but achieving sufficient power levels remains an issue. One colleague has started to use solid state oscillators in CMOS to generate exceptionally high frequencies \cite{Anders:2011if}, since CMOS enables exquisite signal control so that this approach is certainly challenging the use of diodes, but CMOS also suffers from low power output. So what if we miniaturize the gyrotron itself? It will require a concentrated effort, since highly precise 3D micromanufacturing will be required for the miniaturized microwave resonator.


Miniaturization through MEMS technologies make it possible to mass produce technical systems, achieving high precision and robustness in a tiny package, and at a low unit cost. As a result, the patterns of using the systems dramatically change when this becomes possible, because we need to be less careful, an approach which has always benefited research. Another consequence of miniaturization and cost reduction is the ability to parallelize the application. It is therefore interesting to see whether this approach could also benefit NMR, by enabling novel approaches, and new experiments.

\section{Novel applications - the 'killer apps'?}
\begin{figure}[t]
\centering
\includegraphics[width=\columnwidth
]{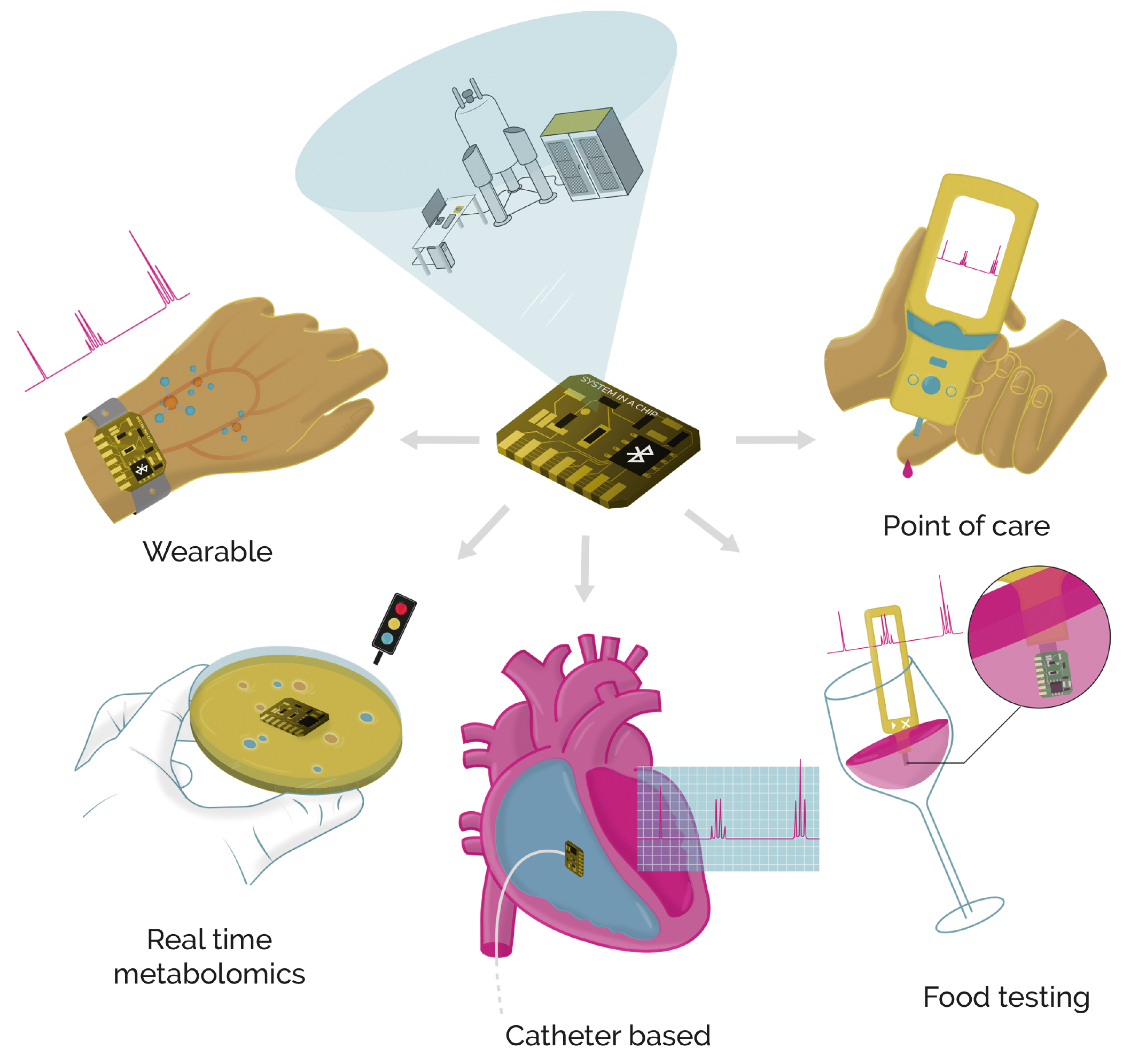}
\caption{Fully integrated chip-based NMR and MRI systems could open up unique applications, from instrumented catheters, to wearable health monitoring systems, point-of care diagnosis, lifestyle assistants, or augmented labware. }
\label{fig:UniqueApplications}
\end{figure}

\begin{figure}[t]
\centering
\includegraphics[width=\columnwidth
]{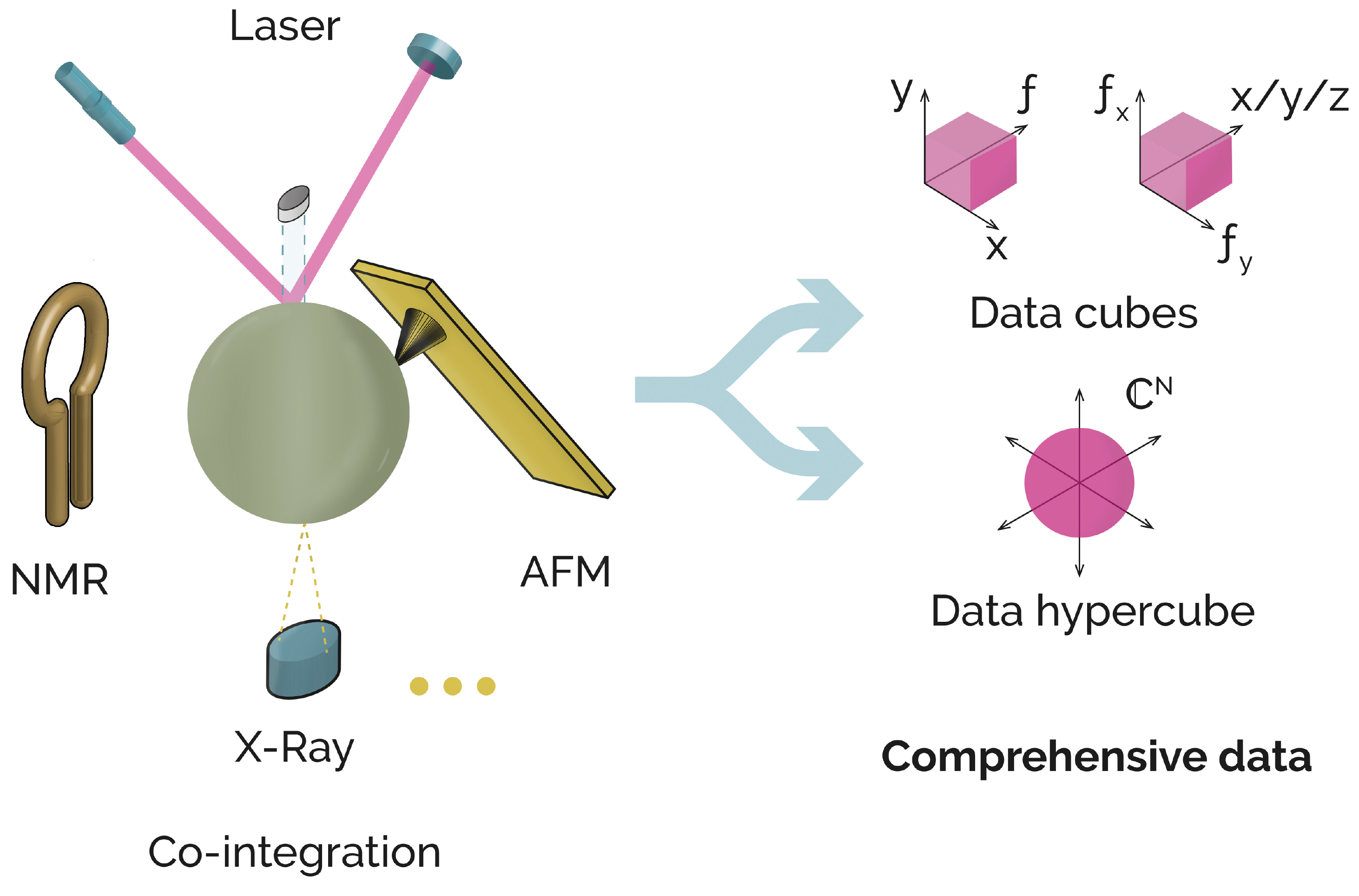}
\caption{Miniaturized NMR would open the opportunity to perform simultaneous measurements across different methods, whilst selectively varying the experimental conditions (excitations) for a single sample. The  acquired data cubes (or hypercubes) would allow the extraction of correlations across all dimensions, to yield more comprehensive information on the material or biological sample, thereby directly informing the creation of digital material twins.}
\label{fig:SumOfItsParts}
\end{figure}

This suggestion is not just speculative. With transistor gates currently at around \SI{10}{\nano\metre}, agile CMOS circuits with billions of transistors capable of gigahertz frequency operation and ultimately low power consumption are routinely manufacturable within a single digit cubic millimetre chip size. The trend in miniaturization is currently (2019) still underway. Further ideas that arise from miniaturisation could be (also see Figure~\ref{fig:UniqueApplications}):

\begin{itemize}
\item \textbf{Disposable sensors.} Current NMR detectors are exquisite instruments,  much like expensive watches, since they are crafted by hand. Miniaturization benefits from precision and automatic parallel manufacturing, so that sensors can become disposable (or perhaps act as sample containers for lengthy storage of rare specimens or forensic evidence).
\item \textbf{Implantable sensors and long term microscopy.} For the extended monitoring of disease progression over time, tiny inductively coupled NMR sensors can be implanted close to the area of interest, offering high enough SNR.
\item \textbf{NMR in a coin; NMR in a pencil.} We should also consider those applications that become possible when NMR is radically miniaturized, i.e., when the entire system is packed into the space of a pencil tip without loss of analysis quality. Such a simple and feasible idea would open up NMR for a vast range of new modalities, perhaps not all of them for scientific applications, but  then NMR will perhaps become ubiquitous. 
\item \textbf{Integration of NMR into  production processes.} Considering the importance of NMR in the field of materials science, and its specific sensing ability, it is remarkable that the method has not yet become a mainstream online control technique in the production of chemicals. Ubiquitous NMR could dramatically change the fields of chemical production (and of course pharamceuticals, cosmetics, energy technology, etc.), if engineering challenges such as compactness and temperature control can be overcome.
\item \textbf{In situ NMR; in vitro metabolomics.} Miniaturized NMR detectors easily fit inside the equipment of other experiments, making it possible to study dynamic phenomena as they happen, localized within very small compartments, see Figure~\ref{fig:SumOfItsParts}. This generic idea holds many opportunities for materials science and biology. Especially in metabolomics, NMR may one day be able to deliver metabolomic rates within a single cell. Even now, miniaturization is enabling the monitoring of metabolomic waste products downstream of cells and small organisms.
\item \textbf{Correlative materials characterization.} Taking the \textit{in situ} and \textit{in operando} idea one step further, the ability to perform a double (triple, ...) measurement at the same sample locality and time, provides the opportunity to correlate physical effects and gain additional insight into material behaviour, for example along coordinates that are otherwise not accessible to NMR. Additionally, these coordinates (light, plasmons, ...)  have their own excitation modalities, which of course can be suitably combined to explore coupled pathways. All this is known, of course, but miniaturization renders these kind of experiments much easier to realize, and may provide improved control over the experiments.
\item \textbf{Opportunities for autonomous operation and discovery.}
By transforming a miniaturized NMR detector into a complete system equipped with additional detectors, as well as advanced computational capabilities, data exploration and data completion approaches more familiar from the world of automomous systems becomes possible. Arrangements of such detectors into cooperating arrays open up the opportunity to more rapidly explore the available information space and complete the spectroscopic picture of a considered sample.
\end{itemize}

\section{Challenges ahead}

NMR microscopy suffers from a \textit{''measurement gap''} challenge. One of the exceptional features of NMR is its non-invasiveness, but all current efforts in driving up detection sensitivity is challenging this notion. 

For example, the most sensitive detectors available are the SQUID, the nitrogen-vacancy (NV) centre, and the magnetic resonance force microscope (the MRFM is essentially an atomic force microscope). SQUIDs, operated at cryogenic temperatures, must be well-insulated from any live matter such as cells, which is hard to do when the systems are miniaturized and brought into close proximity, requiring additional spin-detector distance due to insulation.  NV-centres in diamond, and the MRFM microbeam tip, need to be nanometrically close to target spins, so that these systems cannot operate noninvasively for example in the context of single cell or small organism work. At the other end of the gap we have miniaturized inductive detectors, such as the microcoil or stripline, both of which are sensitive across the diameter of a single cell but require at least $10^{13}$ room temperature spins at 11.74 T for a detectable signal \cite{cho1988nuclear,mcfarland1992three,Olson:1995fr,peck1995design,Massin:2002wu,Maguire:2007ko,Kentgens2008}. 

Taken together, therefore, spatial resolutions between 10 nm and 10 $\mu$m are not adequately covered by sensitive enough non-invasive detectors -- those are three invisible orders of magnitude in distance! The problem is caused by the rapid decay $1/r^3$ over distance $r$ of the dipole strength, which decreases by a factor of $10^{9}$ when going from 10 nm to 10 $\mu$m. At the moment, no principle is in sight for a \textit{''measurement gap''} NMR micro-detector that is sensitive to the small signals produced by typical levels of thermal equilibrium spin polarization in diluted samples. 

From the current perspective, a concerted effort is required. All available SNR must be commandeered of course, but we also have to open up novel hyperpolarization pathways into small organisms or cells, to raise NMR contrast in these very small compartment systems in a gentle manner without disturbing the underlying biological function. Ideally, however, novel detector principles must be devised that reach beyond current inductive detection sensitivity to bridge the \textit{''measurement gap''}.

\section{Acknowledgements} We sincerely acknowledge our collaborators with whom we often speculate about the future of miniaturized NMR. These include from our own group:  J\"urgen Brandner, Lorenzo Bordonali, Pedro Silva, and Dario Mager. Outside of our group we sincerely acknowledge: Jens Anders, Bernhard Bl\"umich, Michael Bock, J\"urgen Hennig, Marcel Utz, and Ulrike Wallrabe. We acknowledge our generous funders, including the European Research Council under Advance Grant number 290586 NMCEL, the European Union's Future and Emerging Technologies Framework (H2020-FETOPEN-1-2016-2017-737043-TISuMR), and the DFG project screeMR (DFG KO 1883/29-1) and Bio-PRICE (DFG BA 4275/4-1, DFG MA 6653/1-1).

\section{References}
\bibliography{references}{}

\begin{thebibliography}{10}

\bibitem{Olson:1995fr}
D.~L. Olson, T.~L. Peck, A.~G. Webb, R.~L. Magin, and J.~V. Sweedler,
  ``{High-Resolution Microcoil 1H-NMR for Mass-Limited, Nanoliter-Volume
  Samples},'' {\em Science}, vol.~270, pp.~1967--1970, Dec. 1995.

\bibitem{Massin:2002wu}
C.~Massin, G.~Boero, F.~Vincent, J.~Abenhaim, P.-A. Besse, and R.~S. Popovic,
  ``{High-Q factor RF planar microcoils for micro-scale NMR spectroscopy},''
  {\em Sensors and Actuators A: Physical}, vol.~97, pp.~280--288, 2002.

\bibitem{Rugar:2004bc}
D.~Rugar, R.~Budakian, H.~J. Mamin, and B.~W. Chui, ``{Single spin detection by
  magnetic resonance force microscopy},'' {\em Nature}, vol.~430, pp.~329 EP--,
  July 2004.

\bibitem{Bentum2007}
P.~van Bentum, J.~Janssen, A.~Kentgens, J.~Bart, and J.~Gardeniers,
  ``{Stripline probes for nuclear magnetic resonance},'' {\em Journal of
  Magnetic Resonance}, vol.~189, pp.~104--113, 2007.

\bibitem{Sakellariou2007}
D.~Sakellariou, G.~L. Goff, and J.~F. Jacquinot, ``{High-resolution,
  high-sensitivity NMR of nanolitre anisotropic samples by coil spinning},''
  {\em Nature}, vol.~447, no.~7145, pp.~694--697, 2007.

\bibitem{Maguire:2007ko}
Y.~Maguire, I.~L. Chuang, S.~Zhang, and N.~Gershenfeld, ``{Ultra-small-sample
  molecular structure detection using microslot waveguide nuclear spin
  resonance},'' {\em Proceedings of the National Academy of Sciences of the
  United States of America}, vol.~104, pp.~9198--9203, May 2007.

\bibitem{Kratt2010}
K.~Kratt, V.~Badilita, T.~Burger, J.~G. Korvink, and U.~Wallrabe, ``{A fully
  MEMS-compatible process for 3D high aspect ratio micro coils obtained with an
  automatic wire bonder},'' {\em Journal of micromechanics and
  microengineering}, vol.~20, no.~1, 2010.

\bibitem{Danieli:2010ep}
E.~Danieli, J.~Perlo, B.~Bl{\"u}mich, and F.~Casanova, ``{Small Magnets for
  Portable NMR Spectrometers},'' {\em Angewandte Chemie International Edition},
  vol.~49, pp.~4133--4135, May 2010.

\bibitem{Anders:2011if}
J.~Anders, P.~Sangiorgio, and G.~Boero, ``{A fully integrated IQ-receiver for
  NMR microscopy},'' {\em Journal of Magnetic Resonance (1969)}, vol.~209,
  pp.~1--7, Mar. 2011.

\bibitem{Ryan2012}
H.~Ryan, S.~H. Song, A.~Za{\ss}, J.~Korvink, and M.~Utz, ``{Contactless NMR
  spectroscopy on a chip},'' {\em Analytical Chemistry}, vol.~84, no.~8,
  pp.~3696--3702, 2012.

\bibitem{WRACHTRUP2016225}
J.~Wrachtrup and A.~Finkler, ``Single spin magnetic resonance,'' {\em Journal
  of Magnetic Resonance}, vol.~269, pp.~225 -- 236, 2016.

\bibitem{Anders:2018}
J.~Anders and J.~G. Korvink, eds., {\em {Micro and Nano Scale NMR: Technologies
  and Systems}}.
\newblock Wiley-VCH, 2018.

\bibitem{PhysRevLett.70.3506}
J.~A. Sidles and D.~Rugar, ``Signal-to-noise ratios in inductive and mechanical
  detection of magnetic resonance,'' {\em Phys. Rev. Lett.}, vol.~70,
  pp.~3506--3509, May 1993.

\bibitem{hoult1976signal}
D.~I. Hoult and R.~Richards, ``The signal-to-noise ratio of the nuclear
  magnetic resonance experiment,'' {\em Journal of Magnetic Resonance (1969)},
  vol.~24, no.~1, pp.~71--85, 1976.

\bibitem{hoult1979sensitivity}
D.~Hoult and P.~C. Lauterbur, ``The sensitivity of the zeugmatographic
  experiment involving human samples,'' {\em Journal of Magnetic Resonance
  (1969)}, vol.~34, no.~2, pp.~425--433, 1979.

\bibitem{kelly2002low}
A.~E. Kelly, H.~D. Ou, R.~Withers, and V.~D{\"o}tsch, ``Low-conductivity
  buffers for high-sensitivity nmr measurements,'' {\em Journal of the American
  Chemical Society}, vol.~124, no.~40, pp.~12013--12019, 2002.

\bibitem{kraus1973electromagnetics}
J.~D. Kraus and K.~R. Carver, {\em Electromagnetics}.
\newblock McGraw-Hill, 1973.

\bibitem{cho1988nuclear}
Z.~Cho, C.~Ahn, S.~Juh, H.~Lee, R.~Jacobs, S.~Lee, J.~Yi, and J.~Jo, ``Nuclear
  magnetic resonance microscopy with 4-$\mu$m resolution: Theoretical study and
  experimental results,'' {\em Medical Physics}, vol.~15, no.~6, pp.~815--824,
  1988.

\bibitem{mcfarland1992three}
E.~McFarland and A.~Mortara, ``Three-dimensional nmr microscopy: improving snr
  with temperature and microcoils,'' {\em Magnetic resonance imaging}, vol.~10,
  no.~2, pp.~279--288, 1992.

\bibitem{Hoult:1976dw}
D.~I. Hoult and R.~E. Richards, ``{The signal-to-noise ratio of the nuclear
  magnetic resonance experiment},'' {\em Journal of Magnetic Resonance (1969)},
  vol.~24, pp.~71--85, Oct. 1976.

\bibitem{C8LC01259H}
L.~Bordonali, N.~Nordin, E.~Fuhrer, N.~MacKinnon, and J.~G. Korvink,
  ``Parahydrogen based {{NMR}} hyperpolarisation goes micro: an alveolus for
  small molecule chemosensing,'' {\em Lab Chip}, vol.~19, pp.~503--512, 2019.

\bibitem{Chen_Magic_2018}
P.~Chen, B.~J. Albert, C.~Gao, N.~Alaniva, L.~E. Price, F.~J. Scott, E.~P.
  Saliba, E.~L. Sesti, P.~T. Judge, E.~W. Fisher, and A.~B. Barnes, ``Magic
  angle spinning spheres,'' {\em Sci Adv}, vol.~4, no.~9, p.~eaau1540, 2018.

\bibitem{Badilita:2012fc}
V.~Badilita, R.~C. Meier, N.~Spengler, U.~Wallrabe, M.~Utz, and J.~G. Korvink,
  ``{Microscale nuclear magnetic resonance: a tool for soft matter research},''
  {\em Soft Matter}, pp.~10583--10597, 2012.

\bibitem{C3RA43758B}
N.~A. Bakhtina and J.~G. Korvink, ``Microfluidic laboratories for {{C.
  elegans}} enhance fundamental studies in biology,'' {\em RSC Adv.}, vol.~4,
  pp.~4691--4709, 2014.

\bibitem{yoon201626}
S.~Yoon, J.~Kim, K.~Cheon, H.~Lee, S.~Hahn, and S.-H. Moon, ``26 {T} 35 mm
  all-{G}d{B}a2{C}u3{O}7--x multi-width no-insulation superconducting magnet,''
  {\em Superconductor Science and Technology}, vol.~29, no.~4, p.~04LT04, 2016.

\bibitem{Kim_Design_2017}
K.~Kim, K.~R. Bhattarai, J.~Jang, Y.~Hwang, K.~Kim, S.~Yoon, S.~Lee, and
  S.~Hahn, ``Design and performance estimation of a 35 {T} 40~mm no-insulation
  {all-REBCO} user magnet,'' {\em Superconductor Science and Technology},
  vol.~30, no.~6, p.~065008, 2017.

\bibitem{Lee:2008bx}
H.~Lee, E.~Sun, D.~Ham, and R.~Weissleder, ``{Chip--NMR biosensor for detection
  and molecular analysis of cells},'' {\em Nature Medicine}, vol.~14,
  pp.~869--874, July 2008.

\bibitem{Takeda:2007cx}
K.~Takeda, ``{A highly integrated FPGA-based nuclear magnetic resonance
  spectrometer},'' {\em Review Of Scientific Instruments}, vol.~78, p.~033103,
  Mar. 2007.

\bibitem{Tang:2011hz}
W.~Tang and W.~Wang, ``{A single-board NMR spectrometer based on a software
  defined radio architecture},'' {\em Measurement Science and Technology},
  vol.~22, no.~1, p.~015902, 2011.

\bibitem{While:ey}
P.~T. While, M.~V. Meissner, and J.~G. Korvink, ``{Insertable biplanar gradient
  coils for magnetic resonance microscopy: theoretical minimization of power
  dissipation for different fabrication methods},'' {\em Biomedical Physics
  {\&} Engineering Express}, vol.~4, no.~3, p.~035019, 2018.

\bibitem{peck1995design}
T.~L. Peck, R.~L. Magin, and P.~C. Lauterbur, ``Design and analysis of
  microcoils for nmr microscopy,'' {\em Journal of Magnetic Resonance, Series
  B}, vol.~108, no.~2, pp.~114--124, 1995.

\bibitem{Kentgens2008}
A.~P. Kentgens, J.~Bart, P.~J. {Van Bentum}, A.~Brinkmann, E.~R. {Van Eck},
  J.~G. Gardeniers, J.~W. Janssen, P.~Knijn, S.~Vasa, and M.~H. Verkuijlen,
  ``{High-resolution liquid- and solid-state nuclear magnetic resonance of
  nanoliter sample volumes using microcoil detectors},'' {\em Journal of
  Chemical Physics}, vol.~128, no.~5, 2008.

\end{thebibliography}
\bibliographystyle{ieeetr}

\begin{figure}[t]
\centering
\label{fig:Discussion1}
\end{figure}

\begin{figure}[t]
\centering
\label{fig:Discussion2}
\end{figure}

\begin{figure}[t]
\centering
\label{fig:Discussion3}
\end{figure}

\begin{figure}[t]
\centering
\label{fig:MagnetSystem}
\end{figure}

\end{document}